\documentclass[12pt]{article}

\usepackage{scicite}

\usepackage{times}

\usepackage{amsmath}
\usepackage{amsfonts}
\usepackage{amssymb}
\usepackage{graphicx}% Include figure files
\graphicspath{{./figures/}}
\usepackage{dcolumn}% Align table columns on decimal point
\usepackage{bm}% bold math

\newenvironment{sciabstract}{
\begin{quote} \bf}
{\end{quote}}

\title{Doppler Compensated Cavity For Atom Interferometry}

\author
{Rustin Nourshargh${}^{1}$, Sam Hedges${}^{1}$, Mehdi Langlois${}^{1}$, Kai Bongs${}^{1}$, Michael Holynski${}^{1\ast}$\\
\\
\normalsize{${}^{1}$MUARC, School of Physics and Astronomy, University of Birmingham}\\
\normalsize{Edgbaston, Birmingham, B15 2TT, UK}\\
\normalsize{$^\ast$Corresponding author. E-mail: M.Holynski@bham.ac.uk}
}

\date{}

\begin{document} 

\baselineskip24pt

\maketitle

\begin{sciabstract}
We propose and demonstrate a scheme to enable Doppler compensation within optical cavities for atom interferometry at significantly increased mode diameters. This has the potential to overcome the primary limitations in cavity enhancement for atom interferometry,  circumventing the cavity linewidth limit and enabling mode filtering, power enhancement, and a large beam diameter simultaneously. This approach combines a magnified linear cavity with an intracavity Pockels cell. The Pockels cell introduces a voltage tunable birefringence allowing the cavity mode frequencies to track the Raman lasers as they scan to compensate for gravitationally induced Doppler shifts, removing the dominant limitation of current cavity enhanced systems. A cavity is built to this geometry and shown to simultaneously realize the capability required for Doppler compensation, with a 5.04~mm $1/e^{2}$ diameter beam waist and an enhancement factor of $>$5x at a finesse of 35. Furthermore, this has a tunable Gouy phase, allowing the suppression of higher order spatial modes and the avoidance of regions of instability. This approach can therefore enable enhanced contrast and longer atom interferometry times while also enabling the key features of cavity enhanced atom interferometry, power enhancement and the reduction of aberrations. This is relevant to future reductions in the optical power requirement of quantum technology, or in providing enhanced performance for atom interferometers targeting fundamental science.
\end{sciabstract}

\section*{Introduction}

Light-pulse atom interferometry is a technique employing atomic superposition states as a sensitive probe. Laser pulses act as atomic beam splitters, transferring atoms into a superposition of states where they are left to evolve. Further pulses recombine the superposition, and the final atomic state populations serve as an exquisite record of the environment in which they evolved. The samples, clouds of laser cooled atoms, are interrogated to measure quantities such as gravity, accelerations, and rotations~\cite{Bongs2019,Freier_2016,Barrett2014875,Budker2007,BIZE2019153}. The sensitivity of these instruments is maximized when the interrogation pulses are completed with high fidelity. This requires laser beams with a large diameter compared to the cloud size, high intensities, and aberration-free optical wavefronts with the latter being a leading cause of uncertainty~\cite{Trimeche2017Wavefront,karcher2018WaveFront,Zhou2016WaveFronts,schkolnik2015Wavefront,hogan2011AgisLeo}. Improvements in instrument sensitivity are achieved by increasing the interferometer space-time area. This can be realized through longer interferometry times or increasing the momentum transferred to the atoms. Interferometry time is typically limited by the available free-fall distance~\cite{Zhou2011}, and the thermal expansion of the cloud. Momentum transfer is limited by the atom-optic beam splitter fidelity which depends on the homogeneity of the beam profile, and the pulse parameters which are in practice limited by the available laser power~\cite{Chiow:12}. Using an optical cavity to provide resonant power enhancement and spatial mode filtering of the atom interferometry beams offers considerable promise in addressing these challenges. Existing cavity enhanced atom interferometers have demonstrated the exciting potential of the technique~\cite{Hamilton2015Cavity,Riou_2017}. Challenges implementing large cavity modes in a geometrically stable way and Doppler shift limited interrogation times have so far made it difficult to deliver on all of the advantages. Overcoming these challenges would allow cavity enhancement to fully benefit atom interferometers, and have the potential to provide significant advantages to a wide range of experiments, including within fundamental physics ~\cite{graham2017mid,badurina2019aion, HindsCopeland2019_DarkEnergy, Tino2008_NewtonG, Parker2018_FineStructure, Overstreet2018_EP,Asenbaum2020_EP, Dimonopoulos2008_GravWaves, jaffe2017testing} and practical applications ~\cite{Bongs2019,wu2019gravity, Hinton2017_PortableMOT, bidel2020_absolute_Airborne}.

During the atom interferometry sequence changes in the atomic cloud velocities, due to gravity or photon recoils, result in a Doppler shift of the laser frequencies as seen by the atoms. Accurate control of the interferometer phase requires compensation of this shift by chirping the laser frequency. In this work we consider an atom interferometer employing two-photon Raman transitions on the D$_2$ line of $^{87}$Rb, requiring the frequency difference between the two Raman lasers to be chirped at $\alpha_{chirp}=25.1~\mathrm{MHz} \, \mathrm{s}^{-1}$. However, operation within an optical cavity places restrictions on the laser frequencies, with implications for atom interferometry as outlined in Materials and Methods. Whilst multiple input frequencies separated by the cavity free spectral range $\nu_{FSR}$ (or multiples thereof) can be resonant simultaneously, standard cavity manipulations do not allow the spacing between resonant modes to be adjusted dynamically. As a result the frequencies cannot be simultaneously resonant with the cavity and compensate the Doppler shift at significant free-fall times, limiting the achievable space-time area for cavity enhanced atom interferometry. This has placed limits on the performance of existing cavity enhanced interferometers~\cite{Riou_2017,Alvarez17BWLimit,alvarez2019optical}. An approach for addressing this has been recently demonstrated~\cite{kristensen2020raman}.

In addition, cavity enhanced atom interferometry with conventional two mirror cavities has been limited by small mode diameters, comparable to the millimeter scale of the atom cloud. This results in dephasing as atoms in the cloud sample different Rabi frequencies at different locations across the cavity mode. This causes a corresponding loss of contrast in the atom interferometer output. In free space atom interferometers a beam to cloud size ratio greater than ten is generally used. Realizing large modes in an optical cavity is a challenging but tractable problem. Recent advancements have yielded a marginally stable cavity with a mode diameter of 1~cm~\cite{Riou_2017} but marginal stability causes higher order mode degeneracy preventing effective spatial mode filtering, one of the key benefits sought through cavity enhancement. 

In this work we demonstrate a scheme for enabling Doppler compensation while achieving large mode diameters and the ability to operate in a stable regime to provide spatial mode filtering. This is achieved through the combination of a magnified linear cavity and an intracavity Pockels cell. This is the first example of a scheme for simultaneously overcoming two of the main limitations in cavity enhanced atom interferometry, limited free-fall times and small mode volumes. 

\begin{figure}[ht!]
	\centering
	\includegraphics[width=.8\columnwidth ]{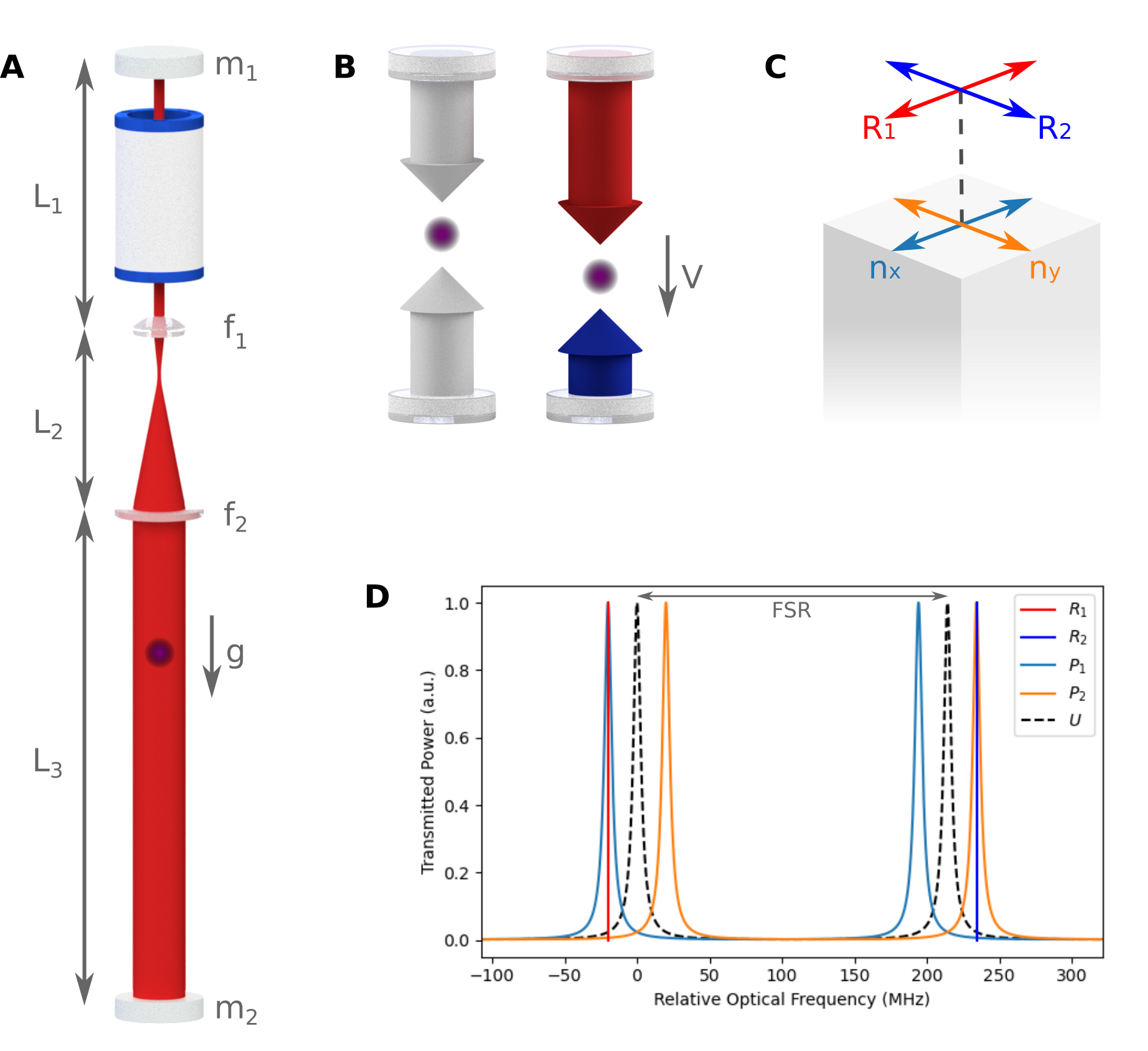}
	\caption{(A): Schematic overview of the cavity, not to scale. The large mode interaction region is the length $L_3$. The small waist and in-coupling mode is on mirror $m_1$ which is planar, as is $m_2$. The Pockels cell is within $L_1$. The two lenses have focal lengths $f_1=12~\mathrm{mm}$ and $f_2=150~\mathrm{mm}$ yielding 12.5$\times$ magnification. (B): atoms (purple) experience Doppler shifted laser frequencies when moving with respect to the cavity. (C): The linearly polarized Raman beams, $R_1$ and $R_2$ are aligned with the orthogonal principle axes of the Pockels cell, $n_x$ and $n_y$ respectively. (D): The Pockels effect enabling tracking of a 40~MHz frequency shift in a simulated cavity with a finesse of 35 and $\nu_{FSR}$ of 217~MHz. $P_1$ and $P_2$ are the resonant frequencies of the two polarizations within the cavity, which track the chirping Raman lasers $R_1$ and $R_2$ in frequency as they compensate for the freefall Doppler shift. U is the uncompensated resonant frequency.}
	\label{fig:CavitySchematic}
\end{figure}

\section*{Results}
Here the magnified geometry and intracavity Pockels cell approach is described and demonstrated. A $12.5\times$ magnification intracavity telescope produces a Gaussian beam with a $1/e^{2}$ diameter of $5.04~\mathrm{mm}$ in a geometrically stable configuration whilst also enabling Gouy phase tuning to maximize higher order mode suppression. An intracavity Pockels cell induces voltage controlled birefringence, lifting the degeneracy of orthogonally polarized cavity modes and enabling independent tuning of their resonant frequencies. Agile control of the Pockels cell birefringence allows the cavity resonances to track the interferometry lasers in frequency as they chirp to compensate for Doppler shifts. 

\subsection*{Mode diameter}
\begin{figure}[ht!]
	\centering
	\includegraphics[width=.8\columnwidth ]{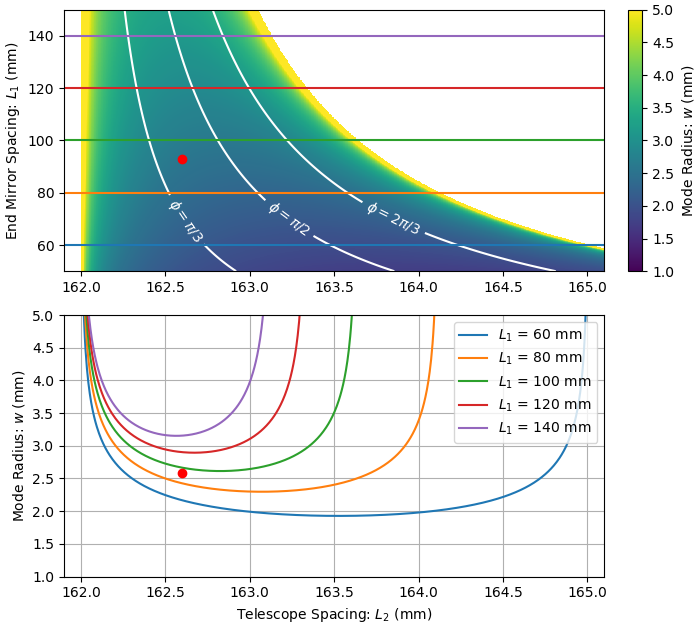}
	\caption{The mode radius in the large diameter section of the cavity is shown for different values of $L_1$ and $L_2$. At the edges of the surface the radius diverges as the cavity reaches the limits of stability ($m\to \pm 1$). Specific slices through the surface illustrate the insensitivity of the mode radius to perturbations in $L_2$. In conventional designs with large modes the behavior is similar to that at the edges of this surface. Owing to the large beam diameter and correspondingly long Rayleigh range, 5.04~mm and 25.6~m respectively, these results are almost completely independent of $L_3$ which is fixed at $L_3 = 0.39$~m. Contours representing the higher order mode degenerate values of Gouy phase are shown in white, and should be avoided to improve spatial mode performance. The cavity described in this work is indicated by the red spot on each plot. The plot was generated by calculating the ABCD matrices for $10^6$ different cavities to select optimal cavity parameters.}
	\label{fig:CavityContour}
\end{figure}

Producing large mode diameters within a cavity is technically challenging, and generally requires either a long cavity, or results in higher order mode degeneracy and marginal stability. We avoid these problems with an intracavity telescope focused near infinity in a two mirror cavity, as shown in figure~\ref{fig:CavitySchematic}. Planar mirrors were selected due to wider availability, leaving five free parameters which must be optimized. Of these $L_2$ and $L_1$ are particularly sensitive. An ABCD-matrix formalism~\cite{Kogelnik66} is used to evaluate the mode radius on mirror-2 ($m_2$) over this parameter space with results shown in figure~\ref{fig:CavityContour}. Optimized parameters are found to be $L_3=390~\mathrm{mm}$ and focal lengths $f_1=12~\mathrm{mm}$ and $f_2=150~\mathrm{mm}$.

We achieve a beam waist of $2w = 5.04~\mathrm{mm}$ on $m_2$, as measured with a CCD. Around the central set of lens spacings ($L_2$), this waist is remarkably insensitive to perturbations in the telescope length. A beam diameter fluctuation of 1\% would require a change in $L_2$ of more than $100~\mu\mathrm{m}$. The intrinsic stability of this cavity geometry is such that we were able to construct it from standard optomechanics mounted to an aluminum rail.

\subsection*{Spatial mode filtering}
An optical cavity only supports spatial modes with well defined wavefront profiles. By coupling into the fundamental mode of the cavity we are able to exploit this property to filter unwanted spatial modes and achieve flat optical wavefronts. To provide such filtering effectively, the cavity design must avoid frequency degeneracy between higher order modes and the fundamental mode. If this is not avoided parasitic coupling to these higher orders will produce significant wavefront distortion. This can be resolved through tuning the Gouy phase. Higher order spatial modes acquire Gouy phase more rapidly than the fundamental Gaussian mode and have different resonant frequencies. The magnitude of the Gouy phase acquired over a round trip is given by 

\begin{equation}
    \phi_{p,q} = (p+q+1) \cos^{-1}(m)
    \label{eq:GouyPhase}
\end{equation}

\noindent for a higher order mode parameterized by indices $p,q$ (e.g. $TEM_{p,q}$) in terms of the half-trace of the ABCD matrix describing the cavity, $m$~\cite{siegman1986lasers}.

Higher order mode degeneracy will occur when $\phi_{p,q}$ is close to a multiple of $\pi$. Adjusting the spacing of the telescope allows the half-trace $m$, and hence the Gouy phase, to be tuned freely over the full range. The Gouy phase is tuned to a value reducing higher order mode degeneracy, whilst maintaining only moderate sensitivity to perturbations in length. Cavity parameters which result in degeneracy with the first five higher order modes are shown in figure~\ref{fig:CavityContour} (three contours and the two boundaries of the surface). We select $m=0.31$ resulting in $\phi=1.25$ avoiding the degenerate values of $\pi/2$ and $\pi/3$. Equivalent Gouy phase adjustments are possible if the lenses are replaced by curved mirrors, reducing round trip loss but introducing astigmatism~\cite{alvarez2019optical}.

\subsection*{Doppler compensation using the Pockels Effect}
\begin{figure}[ht!]
	\centering
	\includegraphics[width=\columnwidth]{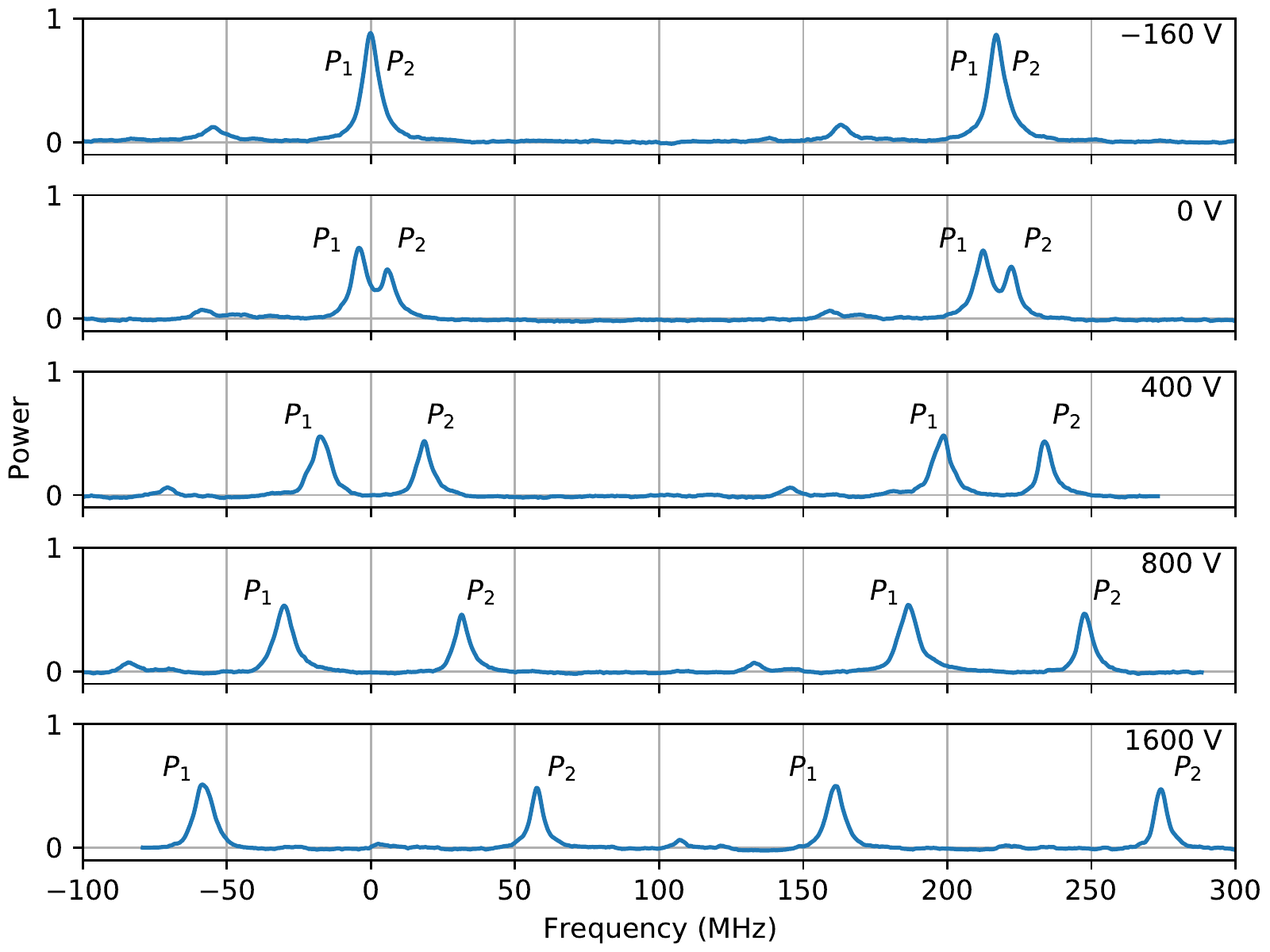}
	\caption{Increasing the voltage applied to the Pockels cell induces birefringence into the cavity. The orthogonally polarized cavity modes are separated in frequency, from the degenerate value at -160~V to 114~MHz at 1600~V applied. The initial offset compensates for residual birefringence in the crystal. Ramping the voltage at $390~\mathrm{V \, s^{-1}}$ compensates for the $25.1~\mathrm{MHz\, s^{-1}}$ Doppler shift.}
	\label{fig:PockelsCellSplit}
\end{figure}

Maintaining cavity resonance whilst independently chirping two input laser frequencies is achieved through the use of an intracavity electro-optic crystal. An intracavity Pockels cell induces a voltage controlled birefringence in the cavity, increasing the optical path length in the slow axis with respect to the fast axis. This lifts the degeneracy between the two orthogonally polarized cavity modes and provides a differential shift of their resonant frequencies allowing them to be independently tuned, see figure~\ref{fig:PockelsCellSplit}. Orthogonal linearly polarized Raman beams implement the atom optics, and the polarization axes are aligned with the principle axes of the Pockels cell, see figure 1C. The frequency shift is linear in applied voltage. This induces a $\lambda /2$ shift in the resonance conditions of $\nu_{FSR}/2$ in a single pass traveling wave cavity and $\nu_{FSR}$ in a double pass standing-wave cavity as presented in this work. The slew rate for Doppler compensation is given by

\begin{equation}
    \frac{dV}{dt}=\frac{2}{N_{passes}}\frac{\alpha_{chirp}}{\nu_{FSR}}V_{\lambda /2}
    \label{eq:SlewRate}
\end{equation}

\noindent where the half-wave voltage depends on the Pockels cell crystal, $V_{\lambda /2} =3.3~\mathrm{kV}$ for our system, and in the standing wave geometry employed here $N_{passes}$ is two. The required slew rate for Doppler compensation of rubidium in free-fall ($\alpha_{chirp} =25.1~\mathrm{MHz~s}^{-1}$), is therefore easily achievable.

Pockels cells are bi-polar, meaning that sweeping the voltage from $-V_{\lambda/2} \to V_{\lambda/2}$ allows the relative frequency to be varied by $\nu_\mathrm{FSR}$ in a traveling wave cavity and $2\nu_\mathrm{FSR}$ in a standing wave cavity. If longer sweeps are required, the resonance condition can be maintained by reversing the polarity of the applied voltage between pulses and noting that the longitudinal modes now in use are separated by an additional free spectral range (in a traveling wave configuration a double length Pockels cell is required to achieve arbitrary frequency shifts).

 An intracavity Pockels cell also relaxes the restrictions on cavity lengths. In previous work, fine tuning of the free spectral range $\nu_{FSR}$ was required to ensure both Raman beams (separated by $\nu_{HFS}$) are resonant with the cavity simultaneously~\cite{Hamilton2015Cavity}. This is satisfied when $\nu_{HFS}= N\nu_{FSR}$, that is, the frequency splitting between Raman beams is an integer multiple of $\nu_{FSR}$. By allowing the initial Pockels cell voltage to take a non-zero value, resonance can be achieved for both Raman beams without restricting $\nu_{FSR}$. In a system where one of the Raman frequencies is generated by electro-optic modulation, having a non-integer $\nu_{FSR}$ separation carries the additional benefit of suppressing any additional frequencies present during the atom interferometry sequence. For instance, for electro-optically generated Raman beams this would suppress unwanted sidebands which would otherwise cause AC-stark shifts and parasitic transitions. Previously such sidebands would have been co-resonant. The degree of suppression will vary as the Pockels cell is scanned and careful choice of pulse timing, cavity $\nu_{FSR}$, and Gouy phase can be used to allow these co-resonances to be avoided.

This technique compensates for Doppler shifts and allows the use of arbitrarily large cavities and at high finesse. The photon lifetime in this cavity is $\sim 50~\mathrm{ns}$, which for typical pulse durations of $1-10~\mathrm{\mu s}$ will not cause significant temporal pulse distortion. For longer, high finesse cavities with a photon lifetime comparable with the pulse duration, distortion of the temporal profile of the pulses becomes significant and further constraints apply~\cite{Riou_2017,Alvarez17BWLimit}.

\subsection*{Power enhancement}
Resonant power enhancement is one of the key benefits we realized with the cavity. This proof of principle system demonstrated a power enhancement of over 5x with a finesse of $\mathcal{F}=35$, using standard components. This has been achieved despite the cavity mode encountering 26 optical surfaces and passing through intracavity optics. Whilst modest for a typical cavity, this is a dramatic enhancement in the context of atom interferometry which could further be improved through optimisation of the optical scheme and components. 

% These improvements combined with the novel design described in this work extend intracavity atom interferometry into a new high power, high fidelity regime, overcoming the central experimental limitations.

\section*{Discussion}
We demonstrate a scheme for cavity enhanced atom interferometry that for the first time combines a large mode diameter with the capability to provide Doppler compensation, reduction of wavefront aberration through spatial mode filtering and power enhancement. This overcomes the two key barriers preventing high performance use of cavity enhanced atom interferometry, while realising the two primary benefits. This is achieved while also being robust to changes in cavity length, avoiding the extreme dimensional tolerances encountered at the edges of geometric stability. The techniques described can enable large-mode, high-finesse cavity experiments capable of compensating for arbitrarily large Doppler shifts due to long free-fall times, and the parameters required to enable large momentum transfer orders. Furthermore, this scheme could be applied to state-of-the-art long baseline atom interferometers, with 10~m baselines and 2~s freefall times~\cite{Dickerson2013}, while still remaining within the limitations imposed by cavity lifetime elongation~\cite{Alvarez17BWLimit}. As such, this scheme is highly applicable to the full range of applications for atom interferometry. For fundamental physics, the scheme could be applied to realize increased optical intensities and reductions in wave-front aberration to enhance space-time area through large momentum transfer without imposing limitations on free-fall time. For quantum technology sensors the scheme could provide reductions in the required laser power, leading to cheaper sensors or operation on power constrained platforms such as satellites, while also providing a route to increased sensitivity through robust cavity enhancement for large momentum transfer.

\section*{Materials and Methods}

\subsection*{Cavity linewidth limit}
\label{sec:LinewidthLimit}

Atoms in an atom interferometer aligned with gravity experience Doppler shifted laser frequencies during freefall. Chirping the lasers to compensate for the resulting Doppler shift has the undesired effect of moving both laser frequencies away from cavity resonance. This cavity linewidth limit results in reduced circulating intensity and prevents beam splitters entirely at long T times. We arbitrarily define a maximum interrogation time $T_{max}$, to be that at which the Doppler shift has reduced the circulating intensity to half of it's resonant value for the final beam splitter pulse ($2T_{max}$ total free fall).

\begin{equation}
    T_{max} = \frac{1}{2}\frac{\nu_{FSR}}{\alpha_{chirp}\mathcal{F}} = \frac{1}{2}\frac{c}{\alpha_{chirp} L_{RT} \mathcal{F}}
    \label{eq:MaxT}
\end{equation}

The free spectral range of the cavity is $\nu_{FSR}$, $\alpha_{chirp}$ is the chirp rate, $\mathcal{F}$ is the finesse, and $L_{RT}$ is the round trip optical path length of the cavity, twice its physical length for standing wave geometries. At $T_{max}$ the Rabi frequency is reduced by a factor of two. Beam splitter pulses at this Doppler shift must be twice as long as on resonance and the velocity class that can be addressed is also reduced by a factor of two. The sensitivity of circulating intensity to perturbations in laser frequency or cavity length are all dramatically increased as the laser moves off of resonance. The first order insensitivity on resonance is replaced by the steep gradient of the Lorentzian Cavity transmission function~\cite{siegman1986lasers}.

\subsection*{Pockels Cell}
The Pockels cell used is a Gooch and Housego Light Gate 3 BBO Pockels cell, AR coated at 780~nm. Precautions should be taken to avoid damage to the crystal caused by the sustained application of high voltages~\cite{GHPockelsAlign}.

\subsection*{Laser system}
Light at 780~nm, resonant with the D2 line of rubidium 87, is generated in fiber by frequency doubling a 1560~nm telecoms laser as shown in figure \ref{fig:InputBb}A. The output from the 1560~nm seed, a Koheras BASIK E15 from NKT, is split into two arms for the two Raman frequencies. One of the arms is also used to stabilize the laser to the cavity using the Pound-Drever-Hall (PDH) technique. In both arms light travels through an electro-optic modulator (EOM), MPZ-LN-10 from iXblue, is amplified by an Erbium-doped fiber amplifier (EDFA), a CEFA-C-PB from Keopsys, frequency doubled by a quasi-phase matched Lithium-Niobate waveguide for second harmonic generation (SHG), WH-0780 from NTT~Electronics, and modulated using an acousto-optic modulator (AOM), MT110-NIR20 from AA~Opto-Electronic, before passing through an OZ Optics optical isolator to a fiber collimator where the light is coupled into free space.

The EOM on the arm not locked to the cavity is modulated at 6.835~GHz; the ground state hyperfine splitting in rubidium 87. To produce this a 7 GHz clock is mixed with 164 MHz from an ARTIQ DDS channel using a single sideband mixer achieving 32.7 dB carrier suppression. The RF signals for the remaining EOM and AOMs are supplied by other ARTIQ DDS channels and an arbitrary function generator.

\subsection*{Locking}
As we inject light into both cavity polarizations, the standard locking scheme using a polarizing beamsplitter and a $\lambda/4$ (quarter-wave plate) to separate the incident and reflected beams to be sampled for PDH stabilization is not viable~\cite{Black01PDH}. Instead we insert a 5\% beam sampler into one of the two polarization beam paths, and stabilize the laser frequency on the sampled fraction of the light reflected from the cavity with standard PDH. This loop is run continuously at low optical power and a large detunig to maintain laser lock whilst minimizing spontaneous emission. The laser is pulsed to full power for the Raman pulses. Adding a sampler and photodiode to the second polarization would allow the Pockels cell voltage to be actively stabilized rather than the existing open-loop configuration.

\subsection*{Cavity mode matching}
\begin{figure}[ht!]
	\centering
	\includegraphics[width=\columnwidth]{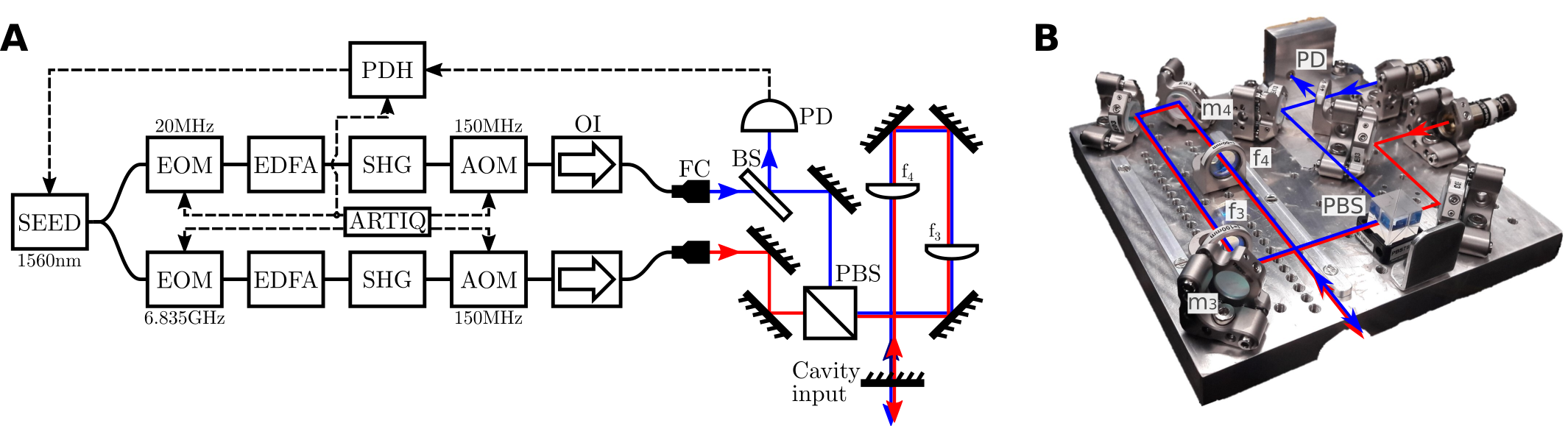}
	\caption{(A): Laser system schematic. Light at 780~nm is produced by amplifying and frequency doubling a 1560~nm seed. (B): Cavity beam delivery board. Mirrors overlap orthogonal linear polarized light from two fiber collimators. The light is reflected through a mode matching telescope and mirrors $m_3$ and $m_4$ are used to achieve spatial overlap with the cavity mode. The cavity reflected light from the polarization shown in blue is reflected by a 5~\% beam sampler onto a bespoke photodiode for locking using the Pound-Drever-Hall technique.}
	\label{fig:InputBb}
\end{figure}

Efficient mode matching is achieved by producing good spatial overlap between the input beam and the cavity mode. We designed a custom beam delivery board to overlap the two laser polarizations on a polarizing beamsplitter, before passing through a telescope, $f_3$ and $f_4$ on figure~\ref{fig:InputBb} to produce a mode of the correct size on the input mirror. Mechanical slides allow the telescope lenses to be translated along the optical axis without introducing tilt or transverse misalignments. The input diameter and location is measured with a CCD which we are able to translate along the cavity rail around the location of the input mirror (removed for this procedure). The telescope lenses are adjusted until the target diameter, determined analytically, is achieved. 

Spatial overlap with the cavity mode is attained with a final mirror pair on the delivery board $m_3$ and $m_4$. These mirrors are adjusted to maximize coupling to the fundamental mode (and minimize coupling to higher orders) as observed on the cavity transmission photodiode (see figure~\ref{fig:PockelsCellSplit}). Precise adjustment of these optics allows mode matching efficiencies above 90\% before adding the Pockels cell.

\subsection*{Optics alignment}
 Despite the large number of optics present in this cavity, optical alignment is relatively straightforward. Beginning with an optic axis parallel to the rail, optics are added in turn and adjusted to overlap back reflections with the input beams, see figure~\ref{fig:CavitySchematic}A. Coarse spacing is measured with mm-level precision, and the critical telescope spacing is adjusted with a micrometer stage to deliver the target beam size from the fixed input, as measured at the waist of the large mode region. Retro reflection from $m_4$ in the absence of $m_3$ allows the orientation of $m_4$ to be precisely defined. $m_3$ is then added, and coarsely aligned by making similar adjustments. Final adjustments should be completed by optimizing the transmission signal. The Pockels cell requires a well centered beam and accurate alignment of roll, tip and tilt, carried out once the cavity is fully aligned. The coarse alignment procedure follows that of standard q-switching applications~\cite{GHPockelsAlign}, whilst the final fine tuning is best achieved by supplying a triangle wave voltage to the cell and adjusting the Pockels cell tip and tilt to minimize voltage induced coupling to higher order modes as observed on the cavity transmission. Temporarily adding a half waveplate on the input to the cavity simplifies precise roll alignment, and enables projection of a single input polarization on both crystal axis for alignment purposes. When the crystal axis is aligned with the cavity axis, the applied voltage induces birefringence, splitting the transmission peak in two and increasing this separation with increased voltage, see figure~\ref{fig:PockelsCellSplit}. The appearance of other peaks or variations in peak amplitude indicate misalignment.

\bibliography{references}
\bibliographystyle{ScienceAdvances}

\noindent \textbf{Acknowledgements:} The authors thank Camron Nourshargh, Farzad Hayati, and Luuk Earl for their contributions.\\
\noindent \textbf{Funding:} This work was supported by the EPSRC under grant number EP/T001046/1 as part of the UK National Quantum Technologies Programme. \\
% \noindent \textbf{Author Contributions:} \\
\noindent \textbf{Competing Interests:} The authors declare that they have no competing interests.\\
\noindent \textbf{Data and materials availability:} Additional data related to this paper may be requested from the authors. \\

\end{document}